\documentstyle[a4,12pt]{article}
\begin{document}
\begin{titlepage} \vspace{0.2in} \begin{flushright}
MITH-98/3 \\ \end{flushright} \vspace*{1.5cm}
\begin{center} {\LARGE \bf  Anomalies in H$_2$O-D$_2$O Mixtures:
Evidence for the Two-Fluid Structure of Water\\} \vspace*{0.8cm}
{\bf M.~Buzzacchi~,~~E.~Del Giudice~~and~~G.~Preparata}\\ \vspace*{1cm}
Dipartimento di Fisica dell'Universit\`a di Milano\\
and I.N.F.N. - Section of Milan\\
Via Celoria 16, 20133 Milan, Italy\\ \vspace*{1.8cm}

{\bf   Abstract  \\ } \end{center} \indent
\baselineskip=18pt

Recent probing of H$_2$O-D$_2$O mixtures by various means (neutron deep 
inelastic scattering, Raman absorption, electrical H$^+$/D$^+$ conductivity) 
revealed an unexpected dependence of the relevant physical quantities on the 
isotopic composition of the mixture. We show that these observations can find
their physical rationale in the context of an 
approach to the physics of liquid water which takes into account the 
non-negligible interaction of the molecules with the electromagnetic field, 
from which a two-fluid microscopical picture of water naturally emerges.
 
\baselineskip=12pt 
\vfill \begin{flushleft}  5 February 1998 \\
\end{flushleft}
\end{titlepage}
\baselineskip=24pt
\section{Introduction}

It is not unusual, and rather fortunate, that a seemingly eccentric research 
programme is vigorously and ably pursued by some scientists, based upon 
motivations that turn out to have poor foundations, but whose outcome unveils 
some subtle and far reaching property of nature. Thus demonstrating once more 
that in science creative error, unlike common-sense prejudice, proves often a 
blessing.

The research programme we have in mind focusses on some remarkably anomalous 
behaviours of H$_2$O-D$_2$O mixtures, when probed by a variety of means: 
deep-inelastic neutron scattering \cite{1}, vibron excitation in Raman 
scattering \cite{2} and H$^+$/D$^+$ conductivity \cite{3}. The original 
motivation of the above experiments \cite{4} is, apparently, the expectation 
that the possible quantum coherence of the pure states (H$_2$O or D$_2$O), or 
Quantum Entanglement (QE), would be disrupted in the mixture of the fluids, 
thus modifying the scattering mechanisms (through the modification of 
quantum interference) at work in the pure liquids. As a result, one would 
thus expect some peculiar dependence on $x_D$ (the molar fraction of D$_2$O in 
the mixture, 0~$\leq~x_{D}~\leq~1$) of the observable cross sections.

As we shall review below, such expectations have been indeed confirmed by the 
experiments [1,2,3]. Have we then found, as the authors of Refs.~[1,2,3] 
claim, a strong evidence for QE~? In a recent comment \cite{5} to the Letter 
of Ref.~[1] we have argued against the conclusions of the authors on the 
basis of two facts:
\begin{enumerate}
\begin{itemize}

\item the large size of the effects: up to 30\% in deep inelastic neutron 
scattering [1];

\item the neutron wavelengths involved in the experiment are smaller than 
0.1$\AA$, more than an order of magnitude smaller than the separations between 
the H~(D) nuclei. As well known, in this physical situation (barring some 
very unusual interaction which, however, is not explicitly mentioned) the 
scattering process is necessarily {\it incoherent}, excluding that any 
spatial coherence in the initial multi-hydrogen wave function may affect the 
scattering process. And the experimental validity of the "impulse 
approximation", based on the incoherence of the scattering process, bears 
witness to the untenability of the idea that QE may explain the stunning 
results of Ref.~[1].
\end{itemize}
\end{enumerate}

In the same comment [5], we also ask the question: if not QE what else could 
explain the anomalous effects in H$_2$O-D$_2$O mixtures? And we argue that 
some good description of the data arises from the assumption that water 
consists of two interpenetrating fluids, one consisting of Coherence Domains 
(CD's), in which the water molecules oscillate in phase with a macroscopic, 
classical configuration of the electromagnetic field trapped within them, while 
the other comprises a dense gas of incoherent molecules that fill the 
interstices among the CD's and is generated by the "evaporation" of the outer 
parts of the CD's due to thermal fluctuations.

The above theory of water has been fully developed and described in Ref.~
\cite{6}, and its foundations in QED have been thoroughly discussed in a 
recent, readily accesible book \cite{7}. Thus, in this paper, we shall only 
briefly recall the main aspect of the new theory of water~[6], based on QED 
coherence, whose validity appears further corroborated, as we shall argue 
in the following, by the anomalous properties of H$_2$O-D$_2$O mixtures.

When looked from a quantum-field theoretical point of view, an ensemble of 
water molecules is a matter quantum field, which in its "Perturbative" Ground 
State (PGS) performs zero-point fluctuations, as dictated by the Heisenberg 
Principle. The same happens to the modes of the quantized electromagnetic 
field. For $T$=0, there exists a critical density $\rho_{C}\simeq0.3$~g/cm$^3$, 
at which the PGS becomes unstable, and the combined system, matter plus 
electromagnetic field, gains energy by condensing a classical electromagnetic 
field, whose oscillations are in phase with those of the matter molecules 
between their molecular ground state and a peculiar excited state, whose 
energy is $E=12.06$~eV. In this way the new ground state, the Coherent Ground 
State (CGS), consists of a highly coherent configuration of matter and 
radiation, which gets successively eroded by the disordering effects of 
temperature, until one reaches the boiling temperature when all molecules 
leave the CGS. It has been shown [6,7] that many of the fundamental 
thermodynamic properties of water can be accurately and naturally described 
by this simple and powerful theory, rigorously based on the fundamental laws 
of Quantum Electrodynamics \footnote{For an independent assessment of the 
latter statement please consult \cite{8}.}. In particular, theory shows that 
the fraction $\xi(T)$ of incoherent molecules as a function of $T$ is given 
by the graph in Fig.1, showing that for $T$=300~K, $\xi\approx$0.7.

But let us see now how these ideas can capture the fascinating physics behind 
the surprising experimental observations of Refs.[1,2,3].

\section{Anomalous neutron deep-inelastic scattering}

An analysis of the anomalous neutron deep-inelastic scattering off nuclei of 
H$_2$O and D$_2$O molecules starting from the two-fluid picture of water that 
QED coherence implies can be performed as follows: due to the phase coherence 
which allows a sharing of the recoil momentum among the large number ($N\simeq
10^{7}$) of molecules clustered in a CD, the coherent fraction is less 
effective in the scattering of neutrons than the incoherent, vapour-like 
fraction.
We can write:
\begin{equation}
\sigma_{H,D}=\sigma_{H,D}^{(i)}\left(\epsilon_{H,D}\frac{N_{H,D}^{(c)}}
{N_{H,D}}+
\frac{N_{H,D}^{(i)}}{N_{H,D}}\right),
\end{equation}
where $N_{H,D}^{(i)}$, $N_{H,D}^{(c)}$ and 
$N_{H,D}=N_{H,D}^{(i)}+N_{H,D}^{(c)}$ are the numbers of incoherent, 
coherent and total H$_2$O and D$_2$O molecules respectively, 
$\sigma_{H,D}^{(i)}$ the deep-inelastic cross sections off the H,~D nuclei of 
the incoherent molecules and $\epsilon_{H,D}<1$ is introduced to account for 
the different cross-sections in the two phases of water.

Due to the two-fluid nature of both H$_2$O and D$_2$O, the fraction 
$\xi_{H,D}=\frac{N_{H,D}^{(i)}}{N_{H,D}}$ of incoherent molecules, will 
in general depend on the 
molar fraction $x_D$ (see below). Anyway, from thermodynamic equilibrium 
one can fix unambiguously the two limiting values:
\begin{equation}
\frac{N_{D}^{(i)}}{N_{D}}\rightarrow 1~~~~~~~  
(x_{D}\rightarrow 0),
\end{equation}
\begin{equation}
\frac{N_{H}^{(i)}}{N_{H}}\rightarrow 1~~~~~~~ 
(x_{D}\rightarrow 1).
\end{equation}
As a result, if 
$Q^{(0)}=\left(\frac{\sigma_{H}}{\sigma_{D}}\right)_{pure}\simeq 10.7$ 
denotes the ratio between the cross sections in the 
pure liquids [1], i.e.
\begin{equation}
Q^{(0)}=\frac{\sigma_{H}^{(i)}}{\sigma_{D}^{(i)}}\frac{\left[\epsilon_{H}\left(
1-\xi_{H}\right)+\xi_{H}\right]}{\left[\epsilon_{D}\left(
1-\xi_{D}\right)+\xi_{D}\right]}
\end{equation}
one obtains the limits:
\begin{equation}
\frac{\sigma_{H}}{\sigma_{D}}\rightarrow Q^{(0)}
\left[\epsilon_{D}\left(
1-\xi_{D}\right)+\xi_{D}\right]~~~~~(x_{D}\rightarrow 0),
\end{equation}
\begin{equation}
\frac{\sigma_{H}}{\sigma_{D}}\rightarrow Q^{(0)}\frac{1}
{\left[\epsilon_{H}\left(
1-\xi_{H}\right)+\xi_{H}\right]}~~~~~(x_{D}\rightarrow 1).
\end{equation}
A satisfactory agreement with the experimental data is obtained by 
choosing $\epsilon_{D}\cong0$, 
$\epsilon_{H}\cong0.5$.
For the intermediate dilution range, let $\eta$ be the ratio between the 
number of incoherent H$_2$O and D$_2$O molecules:
\begin{equation}
N_{H}^{(i)}=\eta N_{D}^{(i)},
\end{equation}
then, assuming for simplicity $\epsilon_{D}=\epsilon_{H}=0$,
\begin{equation}
\frac{\sigma_{H}}{\sigma_{D}}\simeq Q^{(0)}\eta(x_{D})\frac{x_{D}}{1-x_{D}}.
\end{equation}
The experimental dependence of $\frac{\sigma_{H}}{\sigma_{D}}$ on $x_{D}$ 
(see Fig.2) in 
the range 0.3$\leq x_{D}\leq$0.7 can be easily fitted by a straight line:
\begin{equation}
\left(\frac{\sigma_{H}}{\sigma_{D}}\right)_{exp}\simeq 4~+~6.7~x_{D}:
\end{equation} 
this can be used to derive an estimate for $\eta$ at equal H$_2$O-D$_2$O 
concentration ($x_{D}$=0.5):
\begin{equation}
\eta(x_{D}=0.5)=0.83~.
\end{equation}
This result will be rederived and substantiated in the following sections.

\section{Anomalous Raman absorption cross sections}

The same line of thought can lead to an explanation of the anomalous 
absorption cross sections of the OH and OD stretching modes observed in 
mixtures. The experimental results are reported in Fig.3, where the ratio 
$Q=\frac{\sigma_{OH}}{\sigma_{OD}}$ is plotted for several values of the 
concentration $x_D$.

In order to understand these "unreasonable" results, we shall rely on the 
two-fluid picture of water, described in the Introduction. We need, of course 
a scattering mechanism which is substantially different in the two phases. 
While in the incoherent fluid Raman scattering proceeds "as usual", in the 
coherent phase we notice at least two essential differences: the first is the 
energy gap ($\delta=0.26$ eV per molecule) [6] that separates the initial, coherent 
molecular state from the incoherent one, which will lower the energy of the 
relevant intermediate (incoherent) states excited by the initial photon 
beam; while the second is that the final excited stretching mode, being 
incoherent, will lead to an outgoing photon whose energy is lower by the gap 
$\delta$. This line, however, will be of very difficult detection, due to 
its extremely short lifetime to decay back to the coherent ensemble. So, 
under the hypothesis that only incoherent molecules can absorb the incident 
probing radiation, one has for $x_{D}\simeq0.5$:
\begin{equation}
\sigma_{OH}=\frac{N_{H}^{(i)}(x_{D})}{N_{H}^{(i)}(0)}\sigma_{OH}^{pure}=
\frac{\xi(1-x_{D})-\lambda}{\xi}\sigma_{OH}^{pure}
\end{equation}
and
\begin{equation}
\sigma_{OD}=\frac{N_{D}^{(i)}(x_{D})}{N_{H}^{(i)}(1)}\sigma_{OD}^{pure}=
\frac{\xi~x_{D}+\lambda}{\xi}\sigma_{OD}^{pure},
\end{equation}
where $\lambda$ is defined by the relation (see Eq.10): 
\begin{equation}
\eta(x_{D}=0.5)=\frac{\xi~(1-x_{D})-\lambda}{\xi~x_{D}+\lambda}\simeq0.83
\end{equation}
So one obtains:
\begin{equation}
Q\simeq\frac{\xi~(1-x_{D})-\lambda}{\xi~x_{D}+\lambda}\left(\frac
{\sigma_{OH}}{\sigma_{OD}}\right)_{pure}.
\end{equation}
Using the previous values ($\xi$=0.7; $\lambda$=0.03) and setting $x_D$=0.5 one 
gets:
\begin{equation}
Q_{th}\simeq0.84~Q^{(0)},
\end{equation}
whose agreement with experiment is good.

An even more intriguing outcome of this experiment is shown in Fig.4, where the 
relative deviations:
\begin{equation}
\Delta\sigma_{OD}=\frac{\sigma_{OD}(x_{D})-\sigma_{OD}^{(0)}}
{\sigma_{OD}^{(0)}},~~~~~\Delta\sigma_{OH}=\frac{\sigma_{OH}(x_{D})-
\sigma_{OH}^{(0)}}{\sigma_{OH}^{(0)}}
\end{equation}
are given for various mixtures.
One sees that $\Delta\sigma_{OH}~<~0$, but $\Delta\sigma_{OD}~>~0$. If 
the anomaly in the behaviour of $Q$ were due to the lack of formation of the 
coherent dissipative structures in mixtures [1,2,3,4], one should observe 
$\Delta\sigma_{OH}~<~0$ {\bf and} $\Delta\sigma_{OD}~<~0$, since the 
postulated "defective" cooperativity in H$_2$O-D$_2$O mixtures should affect 
both species. An interpretation of this effect is possible if one thinks 
that the scattered intensities are not strictly proportional to the number of 
OH and OD oscillators. Indeed, the two-fluid structure of water implies that, 
since only the incoherent molecules are involved in the interaction with 
the external electromagnetic probe, $\Delta\sigma_{OD}~>~0$ means that 
$N_{D_{2}O}^{(i)}>\xi~x_{D}$ and $\Delta\sigma_{OH}~<~0$ means that 
$N_{H_{2}O}^{(i)}<\xi~(1-x_{D})$. Infact, one has:
\begin{equation}
\Delta\sigma_{OD}(0.5)\simeq\frac{\lambda}{\xi~x_{D}}=+0.09,~~~~~~
\Delta\sigma_{OH}(0.5)\simeq\frac{-\lambda}{\xi~(1-x_{D})}=-0.09,
\end{equation}
which compare well with the results of Fig.4.

In the next section, where we analyse the anomalous H$^{+}$/D$^{+}$ 
conductivity in H$_2$O-D$_2$O mixtures, this effect will be clarified and 
its relation to the two-fluid structure of water physically further 
motivated.

\section{Anomalous H$^{+}$/D$^{+}$ conductivity}

An earlier experiment [3] was performed in order to test the idea that 
thermal fluctuations could induce a continuous formation of short-lived 
coherent structures. According to this view, the non-factorizability of the 
wave wave function of the particles involved in such processes could result 
in the quantum delocalization of H$^+$(OH$^-$) ions and their tunnelling 
through clusters of water molecules (H$_2$O)$_{n}$, thus enhancing the ionic 
conductivity. The chance of formation of these coherent dissipative structures 
would be largely suppressed if we replaced some fraction of the H$_2$O 
molecules with D$_2$O molecules, as in H$_2$O-D$_2$O mixtures, since the 
most favourable environment is clearly one of identical molecules. As a 
consequence, one should observe a decrease in the ionic H$^+$ conductivity 
of mixtures.

This was indeed confirmed experimentally, as we shall briefly review: the 
conductivity of D$^+$ (from DCl) in pure D$_2$O and that of H$^+$ (from HCl) 
in pure H$_2$O were first measured \footnote{The data can be extrapolated to 
obtain the ionic conductivity in the limit of zero ionic concentration}. 
Subsequently, samples of HCl-H$_2$O and DCl-D$_2$O with equal molarity were 
mixed at various $x_D$ concentrations. The experiment measured the combined 
H$^+$/D$^+$ conducitvity $\Lambda$ of the mixtures. Standard electrochemistry 
predicts that $\Lambda$ should depend linearly on $x_D$:
\begin{equation}
\Lambda=\Lambda_{D^{+}}x_{D}+\Lambda_{H^{+}}(1-x_{D}),
\end{equation}
where $\Lambda_{D^{+}}$=312.7~S~cm$^2$~mol$^{-1}$ and 
 $\Lambda_{H^{+}}$=426.3~S~cm$^2$~mol$^{-1}$ are the  
conductivities of the pure samples[3].

The data revealed an unambiguous deviation from (18), with a maximum deviation 
-5.1~\% at $x_D$=0.5~.~~K$^+$ conductivity in H$_2$O-D$_2$O mixtures was 
also measured and only a tiny deviation from linearity (-0.8~\% at $x_D$=0.5) 
was observed. These fact gave substance to the expectation that the above 
mentioned quantum effects might play a role in H$^+$ dynamics in water.

In the following we shall argue that the analysis of water presented in 
Refs.[6,7] provides an alternative explanation of the above effect. As we 
sketched in the Introduction, pure normal and heavy water at thermal 
equilibrium ($T$=300~K) comprise a coherent and an incoherent fraction of 
molecules with relative populations $(1-\xi)\simeq0.3$ and $\xi\simeq0.7$
\footnote{We shall neglect small differences in the QED-coherence properties 
of H$_2$O and D$_2$O.} The stability of the "islands" of coherent matter is 
ensured by the formation of an energy gap, while matter and radiation in the 
incoherent phase retain their uncoupled, perturbative dynamics, which 
renders the incoherent molecules a very dense vapour.

Now, what will happen when we mix H$_2$O and D$_2$O~? A thermodynamical 
argument guarantees that a slight "unbalancement" in the coherent (incoherent) 
H$_2$O populations with respect to $1-\xi$~~($\xi$) in pure H$_2$O must 
occur in the mixture. The same holds for D$_2$O, but the unbalancement is 
in the opposite direction. For thermodynamical equilibrium demands the 
transfer of some D$_2$O coherent molecules into the incoherent phase, while 
an equal number of H$_2$O incoherent molecules gets reabsorbed in the coherent 
phase. Indeed, let us approximate the partition function for incoherent 
molecules:
\begin{equation}
Z_{i}=\frac{V}{N}\frac{T^{3}}{\pi^{5/2}}m^{3/2}(I_{1}I_{2}I_{3})^{1/2},
\end{equation}
where $m$ denotes the mass and the $I_{k}$'s the principal axes of inertia 
of a molecule. If we ignore rotational degrees of freedom (water molecules 
perform only hindered rotations), we have:
\begin{equation}
\frac
{Z_{i}^{D_{2}O}}
{Z_{i}^{H_{2}O}}
\simeq\left(\frac{m_{D_{2}O}}
{m_{H_{2}O}}\right)^{3/2}=1.17,
\end{equation}
while coherence guarantees that the partition functions for 
coherent H$_2$O and D$_2$O molecules are equal:
\begin{equation}
\frac
{Z_{c}^{D_{2}O}}
{Z_{c}^{H_{2}O}}=1.
\end{equation}
Equilibrium is reached when the chemical potential is constant 
everywhere and this can be 
achieved through the described transfer mechanism: at equal H$_2$O-D$_2$O 
concentration ($x_D=$0.5) the incoherent fraction gets enriched in heavy 
water by about 15\%. With this in mind, and the consideration that the 
scattering processes responsible for resisitivity (or its inverse, 
conductivity) take place essentially within the incoherent fraction, we can 
provide a different picture of the anomalous proton/deuteron conductivity 
in mixtures: as a first approximation one has for the scattering cross sections:
\begin{equation}
\sigma_{D}(x_{D})=x_{D}\sigma_{D,D_{2}O}+(1-x_{D})\sigma_{D,H_{2}O},
\end{equation}
\begin{equation}
\sigma_{H}(x_{D})=x_{D}\sigma_{H,D_{2}O}+(1-x_{D})\sigma_{H,H_{2}O}.
\end{equation}
This holds when the number of H$_2$O and D$_2$O molecules involved in the 
scattering are exactly proportional to their concentrations, but according to 
the previous considerations one has:
\begin{equation}
N_{H_{2}O}^{(i)}=[(1-x_{D})\xi-\lambda]N,~~~~~~~
N_{D_{2}O}^{(i)}=(x_{D}\xi+\lambda)N,
\end{equation}
where $\lambda$ has the value~0.03. As a consequence, at least for dilutions 
not too different from 0.5, one must change (22),(23) into:
\begin{equation}
\sigma_{D}(x_{D})=\frac{\xi x_{D}+\lambda}{\xi}\sigma_{D,D_{2}O}+
\frac{(1-x_{D})\xi-\lambda}{\xi}\sigma_{D,H_{2}O},
\end{equation}
\begin{equation}
\sigma_{H}(x_{D})=\frac{\xi x_{D}+\lambda}{\xi}\sigma_{H,D_{2}O}+
\frac{(1-x_{D})\xi-\lambda}{\xi}\sigma_{H,H_{2}O},
\end{equation}
and we choose $\sigma_{D,D_{2}O}=1.364$~[3], $\sigma_{D,H_{2}O}=
\sigma_{H,D_{2}O}\simeq1.364^{1/2}$ relative to $\sigma_{H,H_{2}O}=1$~[3].
The conductivity of the mixtures is then given by:
\begin{equation}
\Lambda(x_{D})=x_{D}\frac{\sigma_{D}(x_{D}=1)}{\sigma_{D}(x_{D})}\Lambda_{D}+
(1-x_{D})\frac{\sigma_{H}(x_{D}=0)}{\sigma_{H}(x_{D})}\Lambda_{H},
\end{equation}
whose behaviour as a function of $x_{D}$ is plotted in Fig.5.
This simple calculation, which is a rather straightforward consequence of the 
ideas developed in Refs. [6,7] gives a remarkable agreement with the 
experimental data [3], thus providing a further corroboration of the two-
fluid structure of water.

\section{Conclusions}

The main aim of this paper, as explained in the Introduction, was two-fold. 
On one hand to emphasize the importance of the experimental 
observations of unexpected effects in H$_2$O-D$_2$O mixtures~[1,2,3] 
for our understanding of a fundamental physical system, such as water; and
on the other to show that their explanation does not involve the rather 
arcane (and definitely "untenable") mechanisms of Quantum Entanglement (QE) 
but provides a strong and convincing evidence of the two-fluid nature of 
water, and of the theory of QED coherence [6,7,8] which lies at its roots.

We believe (and the unconvinced reader is invited to strongly object) 
that both goals have been attained. For we have brought to 
focus, and together, three "anomalies" of the H$_2$O-D$_2$O 
mixtures that should (we don't know whether they will) cause more than one 
sleepless night to the molecular dynamicists who assert that through 
Montecarlo simulations they understand water, both light and heavy.
In addition we have argued, we hope convincingly, that at least for deep-
inelastic neutron scattering [1] and the vibron excitation in Raman 
scattering [2] QE cannot be the explanation due to basic theoretical reasons 
for the former and to experimental discrepancies for the latter. On the 
other hand, the simple and well defined two-fluid picture of water, both 
light and heavy, that naturally arises from a theory of condensed matter 
that has finally been able to include among the relevant interactions the 
electrodynamic one (and this in a full quantum field theoretical framework) has 
been shown to remarkably account in a quantitative fashion for the subtle and 
strange phenomena we have analysed in this paper.

In conclusion, the main lesson we think can be learned from the 
present analysis is that, even though the vast majority 
of the physics literature is today devoted to "normal" 
(in the sense of T.~Kuhn) science, where the generally accepted 
"paradigm" cannot but be corroborated in an atmosphere of collective 
accomplishment and self-satisfaction, some unexpected progress may arise from 
the little "anomalies" that get readily accepted on the basis of some more or 
less obscure (and modern quantum mechanics proves prodigious at that) arguments 
that do not seem to jeopardize the "paradigm". And the progress we have in mind 
is the shift from the defective electrostatic paradigm of today's condensed 
matter (the "Electrostatic Meccano", as we like to call it) to electrodynamic 
coherence, that takes full account of the neglected, but very real and deeply 
rooted in QED, long range interaction between matter and the electromagnetic 
field.

How many more far reaching "anomalies" shall we be able to 
observe when this shift will be accomplished~?~~It is a 
question on whose answer we can at present only dream.

\newpage

\begin{center}
FIGURE CAPTIONS
\end{center}
\vskip1.5cm
{\small {\bf Fig.1}: The incoherent fraction of water molecules as a function 
of $T$.}
\vskip1cm
{\small{\bf Fig.2}: (o) Anomalous n-DIS in H$_2$O-D$_2$O mixtures : 
no dependence of $Q$ on $x_D$ was expected; (*) our prediction for $x_D$=0.5}
\vskip1cm
{\small{\bf Fig.3}: (o) Anomalous Raman absorption in H$_2$O-D$_2$O mixtures: 
no dependence of $Q$ on $x_D$ was expected; (*) our prediction for $x_D$=0.5}
\vskip1cm
{\small {\bf Fig.4}: $\Delta\sigma_{OD}$ and $\Delta\sigma_{OH}$ as a function 
of $x_{D}$; (*) our predictions for $x_D$=0.5}
\vskip1cm
{\small{\bf Fig.5}: H$^{+}$/D$^{+}$ conductivity in H$_2$O-D$_2$O mixtures. 
Solid line: our calculation. Circles: experimental data [3]. Dashed line: 
theoretically expected conductivity.}
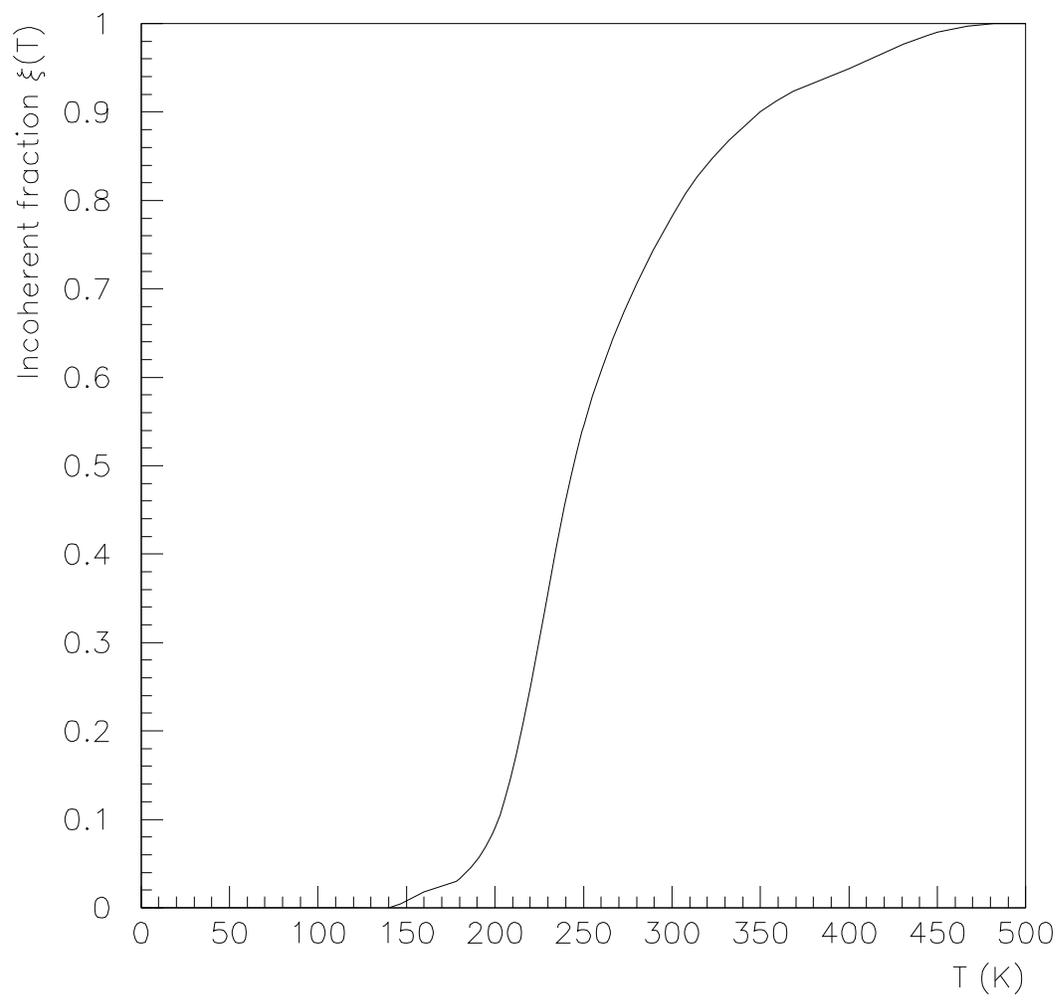
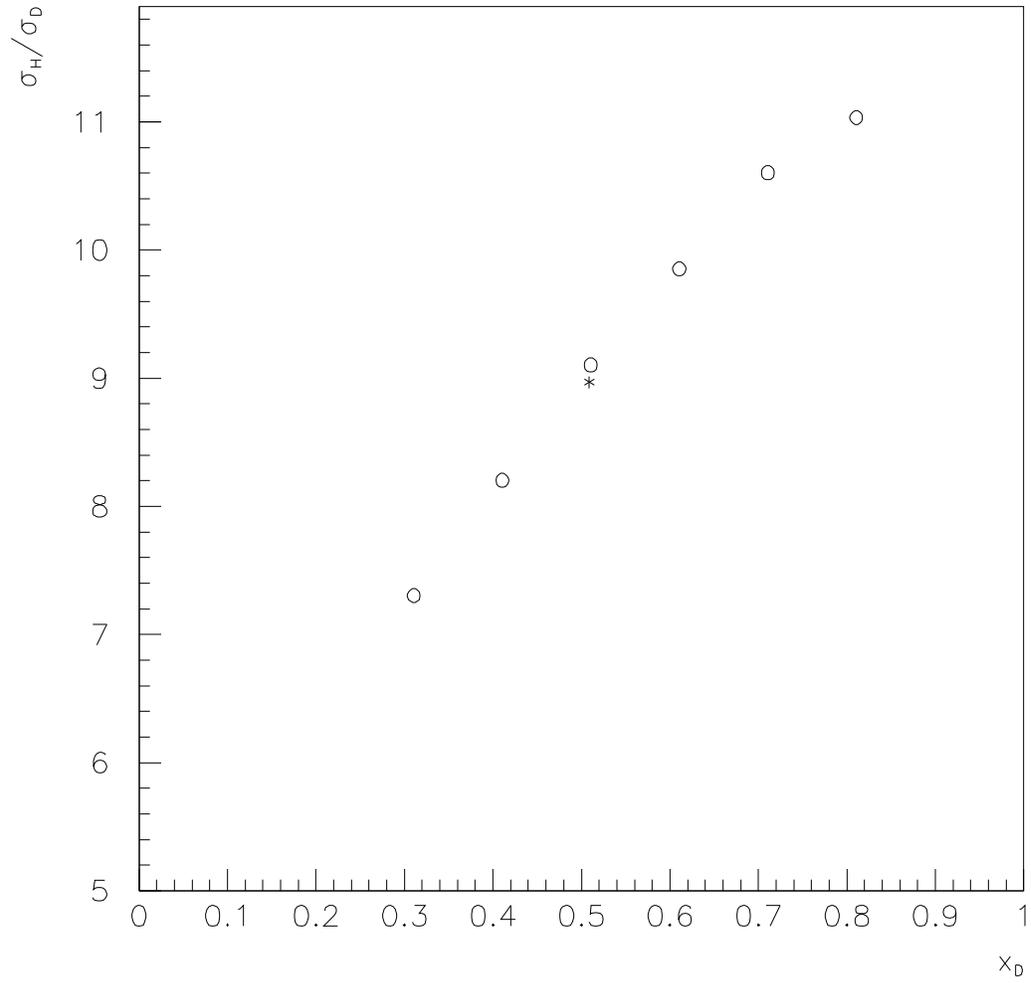
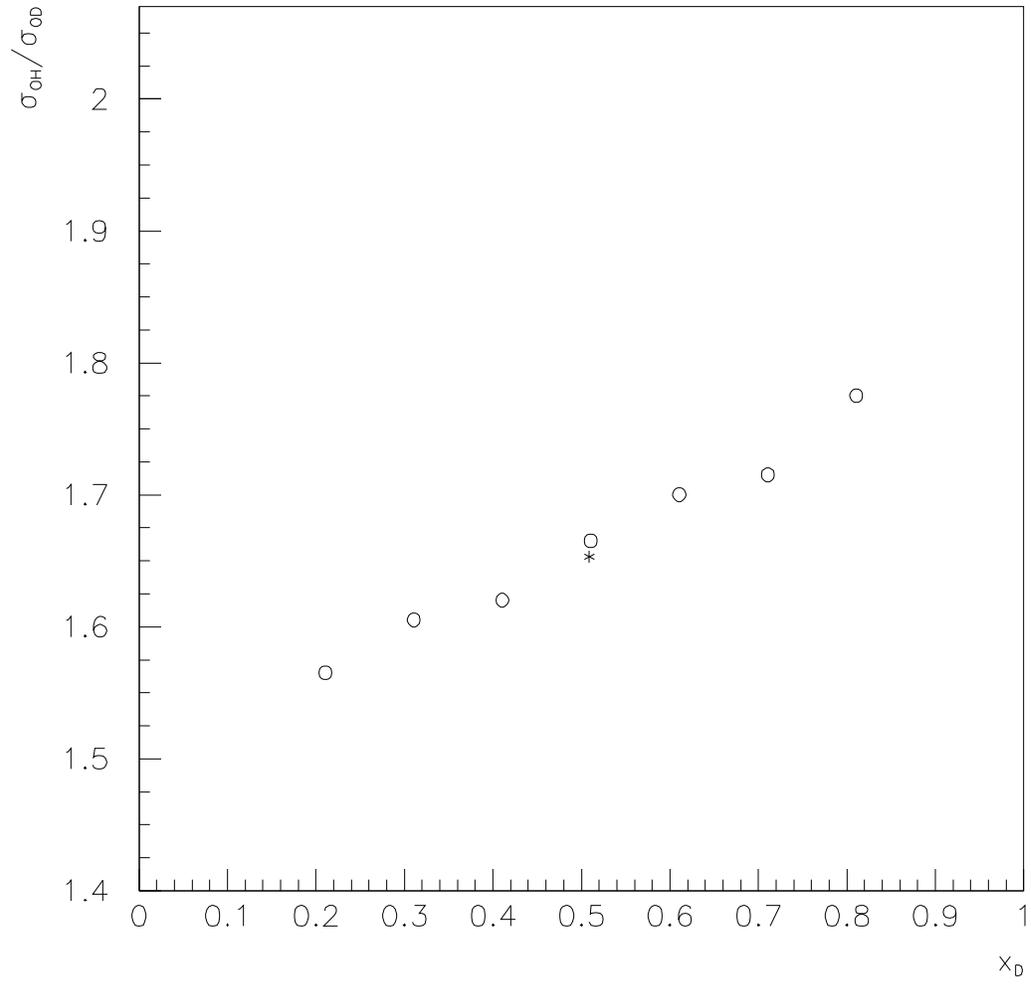
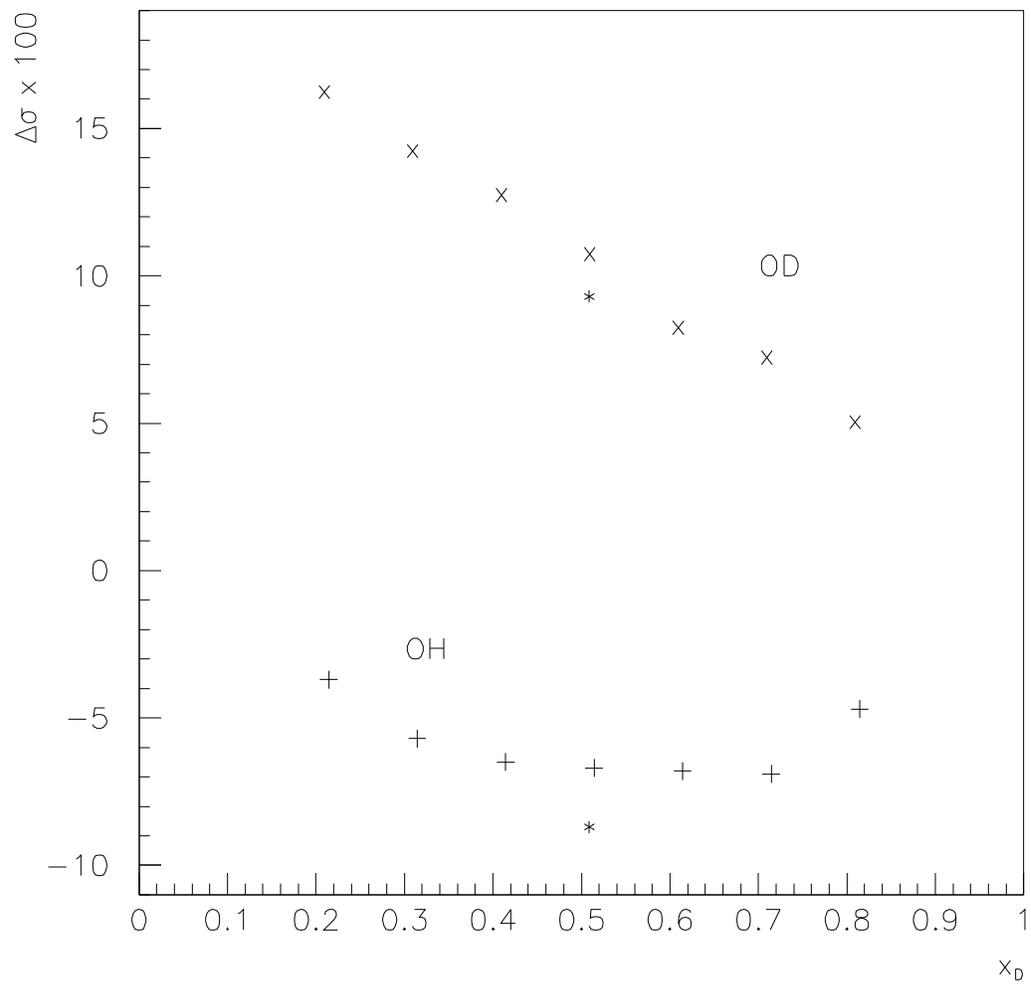
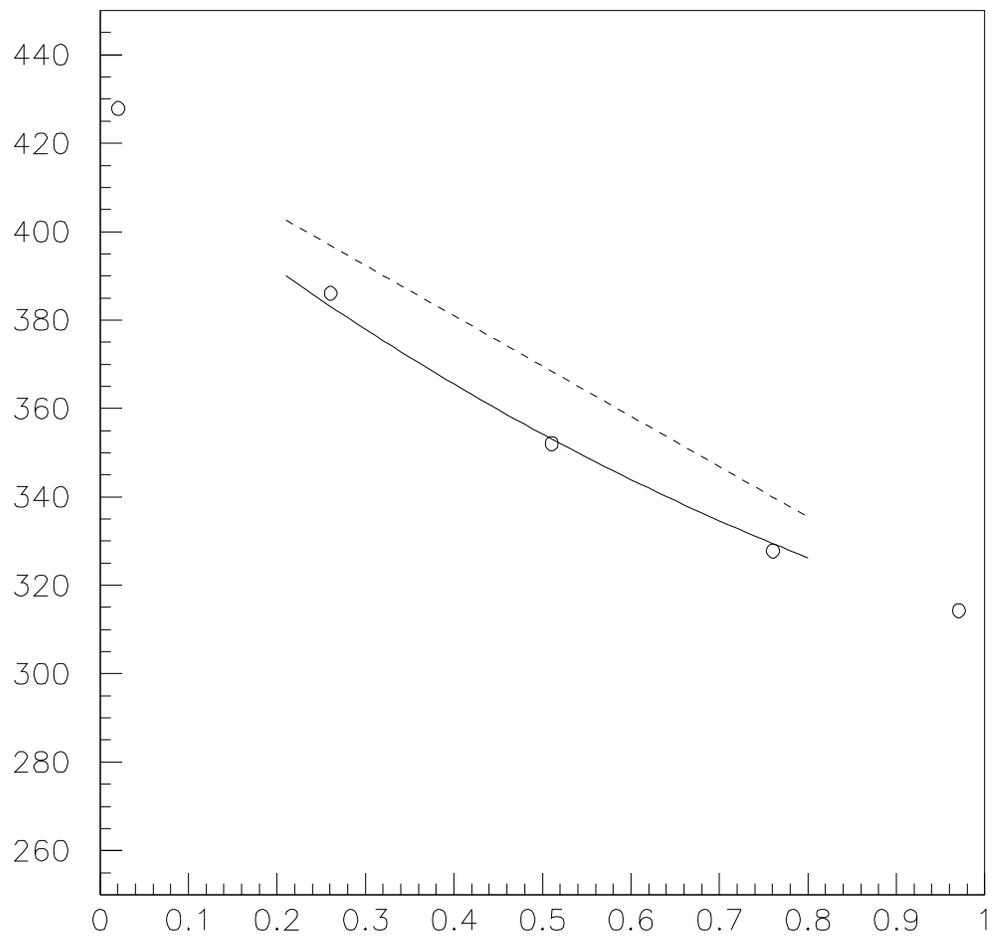
\end{document}